\documentclass{article}

\usepackage[style=numeric, sorting=none, doi=true, maxcitenames=3, mincitenames=1]{biblatex}
\usepackage{authblk}
\usepackage[a4paper, total={6in, 9in}]{geometry}
\usepackage{blindtext}
\usepackage{amsmath, amssymb}
\usepackage{graphicx}
\usepackage{float}
\usepackage{hyperref}
\usepackage{subcaption}
\usepackage[version=4]{mhchem}
\usepackage{multirow}

\renewenvironment{abstract}{%
	\begin{center}
		\begin{minipage}{0.75\textwidth}
			\footnotesize
			\textbf{Abstract: }
		}{%
		\end{minipage}
	\end{center}
	\vspace{5mm}
}

\title{Using Geometric Deep Learning for the Detection of the Diffuse Supernova Neutrino Background in Water-Cherenkov Detectors}

\author[1]{Matthieu Melennec}
\author[1,2]{Andrew D. Santos}
\author[1]{Thomas Mueller}
\author[1]{Fr\'ed\'eric Magniette}

\affil[1]{Ecole Polytechnique, IN2P3-CNRS, Laboratoire Leprince-Ringuet, F-91120 Palaiseau, France}
\affil[2]{Kavli Institute for the Physics and Mathematics of the Universe (WPI), The University of Tokyo Institutes for Advanced Study,	University of Tokyo, Kashiwa, Chiba 277-8583, Japan}

\date{}

\begin{document}
	
	\maketitle
	
	\begin{abstract}
	 	The study of the Diffuse Supernova Neutrino Background (DSNB) is an essential approach to improve our understanding of the formation of Core-Collapse Supernovae, as well as investigating fundamental neutrino properties. The detection of the DSNB flux however is a difficult task, as it requires the disentanglement of its signal with many competitive backgrounds in the low energy range. The main DSNB detection channel is the Inverse Beta Decay (IBD), which is characterised by a distinct, time-separated two-signal signature. This work introduces the use of Graph Neural Networks (GNNs) to handle critical detection and reconstruction tasks for the DSNB through the IBD channel. In particular, we present a model to recover the energy of the first signal from a low energy positron. Additionally, we develop a signal-to-background classifier optimised for the secondary signal, which corresponds to the characteristic photon emission following a neutron capture event. We test our models on two simulations in a Hyper-Kamiokande-like water Cherenkov detector geometry, and show that we obtain performances that are comparable to the state of the art. 
	\end{abstract}
	
	\section{Introduction}
		
		Core-Collapse Supernovae (CCSNe) are caused by the cataclysmic death of massive stars \cite{janka_07}. Their study is essential not only for understanding the CCSN mechanism, but also to study neutrino properties ~\cite{muller_19}. It however faces a major difficulty, which is the frequency of CCSNe in the Milky Way, with roughly a couple per century \cite{rozwadowska_21}. A promising alternative would be the analysis of the Diffuse Supernova Neutrino Background (DSNB) \cite{beacom_10}, which is the background signal consisting of neutrinos from all the core-collapse supernovae that have ever happened in the history of the universe. This background provides a continuous and constant flux of CCSN neutrinos, giving a quality channel for their study.
		
		The DSNB however is a low energy flux and competes with many backgrounds \cite{beacom_10}. These include reactor neutrinos, cosmic muon spallation signals and atmospheric neutrinos. It is therefore necessary to reduce all these backgrounds to efficiently and reliably detect it. The main DSNB detection method in water-Cherenkov detectors, such as the future Hyper-Kamiokande (HK) detector \cite{hk_18}, is through the Inverse Beta Decay (IBD) channel \cite{elhedri_21}. Antineutrino interactions with protons in the water produce a positron, which emits a low energy Cherenkov signal ($[3, 30]$ MeV), and a neutron, which after thermalisation in pure water ($\sim 200 \ \mu$s) is captured by a nucleus, emitting a 2.2 MeV photon. The IBD has the biggest cross-section in the DSNB energy range \cite{giampaolo_23}, and its two-phase signature makes it relatively straightforward to disentangle from noise. The IBD detection reconstructs the features of positron-like signals (called the prompt signal), and looks in a 500 $\mu$s window for a delayed signal, which must be classified as a neutron capture signal or as noise. The most common approach \cite{elhedri_21} regresses prompt signal features (vertex, timing and direction) by fitting signal distributions, and selecting the likeliest set of features \cite{smy_07}. The positron energy is obtained from an effective number of hits around the expected arrival time \cite{abe_11}. The neutron capture identification, or neutron-tagging, is done by extracting several relevant variables from potential neutron capture signals and classifying them as signal or noise based on these features \cite{abe_21}.
		
		The low DSNB flux, as well as the need for the dual detection of prompt and delayed signals imposes a strict computational constraint, as detection algorithms must be particularly efficient, as well as sufficiently performant to detect as many IBD signals as possible, with reduced background contamination \cite{maksimovic_21, li_22}. 
		Machine learning algorithms, and deep learning approaches such as Convolution Neural Networks (CNNs) in particular, have been shown to perform tasks such as Cherenkov-signal reconstruction and signal classification particularly well \cite{maksimovic_21, choma_18,}. 
		However, the non-Euclidean nature of detectors such as the cylindrical Hyper-Kamiokande detector however poses a grave issue, with artificial discontinuities needed to flatten the detector geometry. Furthermore, the sparse nature of signals implies that the computational cost is uselessly increased by applying convolution kernels on regions with no recorded hits. Lastly, water-Cherenkov detectors, such as HK, might have irregular photomultiplier tube (PMT) layouts, with the use of smaller multi-PMT clusters \cite{hk_18}. Graph Neural Networks (GNNs) provide a solution to these issues. They are a generalisation of image convolution to unordered, variable size and irregular point-cloud structure \cite{kipf_17}. Additionally, they adapt better to the sparse nature of these signals \cite{kipf_17, gilmer_17}. These qualities make them very promising candidates to perform the above tasks within the performance and computational constraints imposed by IBD detection.
		
		The objective of this work is to evaluate whether a common geometric deep-learning framework can be adapted to two complementary reconstruction tasks relevant to IBD detection: prompt positron energy regression and delayed neutron-capture tagging. As such, we first present the data simulation that was used. It uses a geometry of HK to simulate both elements of the IBD signal. We then introduce the building blocks of a GNN, and how a full model can be built from them. This structure is first applied to the regression of the prompt signal's energy. We show that the achieved performance is comparable to the state of the art, and discuss potential further improvements. Lastly, we apply our algorithms to the task of neutron tagging, and propose our model architecture as a baseline for the application of modern machine learning methods to this task in Water-Cherenkov detectors.
				
	\section{Data Simulation}
		
		The detection of the DSNB through the IBD channel works following a two-step approach, regressing the prompt signal features on one side, and tagging a delayed signal as a neutron capture on the other side. As such, we simulate two datasets, one for each of these signals. Simulations are performed using WCSim \cite{wcsim_24}, a generic Water-Cherenkov simulation software based on \textsc{geant4} \cite{agostinelli_03}. We used, in particular, a geometry for Hyper-Kamiokande, the WCSim \texttt{HyperK\_HybridmPMT\_IDonly\_Realistic} geometry, which corresponds to the tank of Hyper-Kamiokande and the inner PMTs of the detector. In this geometry, there are two different PMT types: the first type are a single PMT, while the second are multi-PMT (mPMT), which are clusters of 19 smaller PMTs. The dark noise rate is set to 8.4 kHz for single PMTs, and to 1 kHz for PMT clusters \cite{dealtry_17}.
		
\begin{figure}
			\centering
			\includegraphics[width=0.7\textwidth]{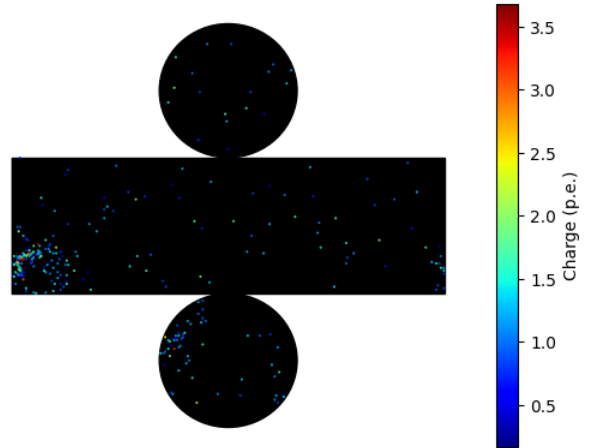}
			\caption{Prompt event in the HK detector. One recognises the circular photon deposition,	typical of Cherenkov cone emission, but it is notably fainter than for high energy events.}\label{sec:data_fig:prompt_evt_display}
		\end{figure}

		For the prompt event dataset, we are interested in simulating the traversal of the detector by a positron. The main background source is the dark noise from the PMTs. As such, we sample positrons with uniformly distributed energy between 3 and 30 MeV, placed uniformly at random in the tank, and with isotropically distributed directions in the detector. We use WCSim's default digitiser and trigger settings for the \texttt{HyperK\_HybridmPMT\_IDonly\_Realistic} geometry. A typical low energy signal deposited in the detector, over a 1 $\mu$s window around the positron's Cherenkov emission, is shown in Figure \ref{sec:data_fig:prompt_evt_display}.
		
		For neutrons, we are looking for neutron capture gamma emission signals, with as most competitive background the dark noise of the PMTs. As such, we use the WCSim software to	simulate a 1 MeV neutron, uniformly at random in the detector, which, after some free path, is captured by a nucleus, emitting a 2.2 MeV photon.  Note that one must not emulate any triggering mechanism, as the 2.2 MeV signal is well below standard triggers. The resulting signal looks very much like a dark noise simulation, and is almost impossible to differentiate with a human eye, and in particular without the further pre-processing presented
		in section \ref{sec:neutron_tagging}.
		
	\section{Algorithmic Framework}
		
		The core of the reconstruction pipeline is based off of Graph  Neural Networks \cite{gilmer_17}. As such, it is essential to build a graph representation of the data that encodes the relevant relationships between individual PMT hits. A graph is a structure $G=\{V, E\}$, where $V$ is the set of nodes $v$, and $E$ the set of edges $e_{vw}$, pairing nodes $v, w\in V$ \cite{bretto_12}. Each node and edge carry a set of features. The selected approach for graph construction from the raw data consists of using the $k$ Nearest Neighbours ($k$-NN) algorithm to construct a graph which encodes the locality of node structures, needed to identify the ring-like structure of Cherenkov signals. To further improve the precision of this representation, we decide to use the timing information, which allows the $k$-NN algorithm to encode both the geometric and timing proximities of hits. As such, we use a Euclidean embedding of the space-time coordinates $(ct, x, y, z)$ (not the Lorentzian interval), since only unit-consistent locality in the k-NN sense matters here.
		The resulting edges carry as features the euclidean space-time distance between their end nodes. To encourage feature flow through the graph, as well as to avoid adding any unphysical bias, we build undirected graphs, i.e. $e_{vw} = e_{wv}$.
		
		The GNN is structured around successive layers of convolution and pooling layers. The convolution operation consists in the collection, by each node $v\in V$, of features obtained from their set of neighbours $\mathcal{N}(v)$. The pooling layers reduce the node count, in order to increase the receptive field of the convolution operation by creating connections between nodes that are further apart. 
		
		Our convolution operation is based on message passing convolution, which computes the updated features $\xi_v$ from the node features $\zeta_v$ and its neighbours as \cite{gilmer_17}
		\begin{equation}
			\xi_v = \gamma_{\Theta_\gamma} \left[ \zeta_v, \underset{w\in \mathcal{N}(v)}{\square} \phi_{\Theta_\phi} \left( \zeta_v, \zeta_w, e_{vw} \right) \right],
		\end{equation}
		where $\phi_{\Theta_\phi}$ is the message function, combining the features of $v$, of a neighbour $w\in \mathcal{N}(v)$ and the features of their linking edge $e_{vw}$. It is parametrised by learnable parameters $\Theta_\phi$, and is usually implemented as a shallow Multi-Layer Perceptron (MLP). $\square$ is the aggregator, updating the current node’s features with the aggregated messages from neighbours. Since the number of neighbours of $v$ may vary, the aggregation operator must be normalised and commutative. Typical aggregators are max or mean functions. $\gamma_{\Theta_\gamma}$ is the update function, mixing the aggregated messages and	the node features. It can be very simple or implemented as a shallow MLP, and has trainable parameters $\Theta_\gamma$. The idea of graph convolution is that each node collects a collection of messages from all of its neighbours. Once all messages are collected, the aggregator makes a synthesis of the messages. This synthesis of its surroundings is what is used by the node to update its features.
		
		Our implementation simplifies the update function by adding self-loops, i.e. considering individual nodes as their own neighbours, which is equivalent to implementing a dedicated update function \cite{gilmer_17, kipf_17}. The message function $\phi_{\Theta_\phi}$ is a MLP with Leaky-ReLU activation function, which takes as input the concatenated features of $v$, its neighbour $w$, and their separating edge $e_{vw}$. The aggregator will be chosen to be either a mean, for regression tasks, or max operation for classification tasks \cite{goodfellow_16}. As such, our convolution operations are based on the convolution function defined as
		\begin{equation}
			\xi_v = \underset{w\in \tilde{\mathcal{N}}(v)}{\square} \phi_{\Theta} \left( \text{concat} \left[ \zeta_v, \zeta_w, e_{vw} \right] \right),
		\end{equation}
		with $\phi_\Theta$ an MLP with trainable parameters $\Theta$.
		
		Graph pooling is an important step in graph convolution, as it allows to coarsen the graph, hence looking at the graph in a more global approach \cite{defferrard_16}. First, let us express the steps of a graph pooling operation as in \cite{grattarola_24}, with a clustering (or selection) step, where we select which nodes will be clustered together, followed by a reduction step, where nodes within clusters are pooled, and finally a connection step, during which we update the adjacency of clusters. 
		
\begin{figure}
			\centering
			\includegraphics[width=0.8\textwidth]{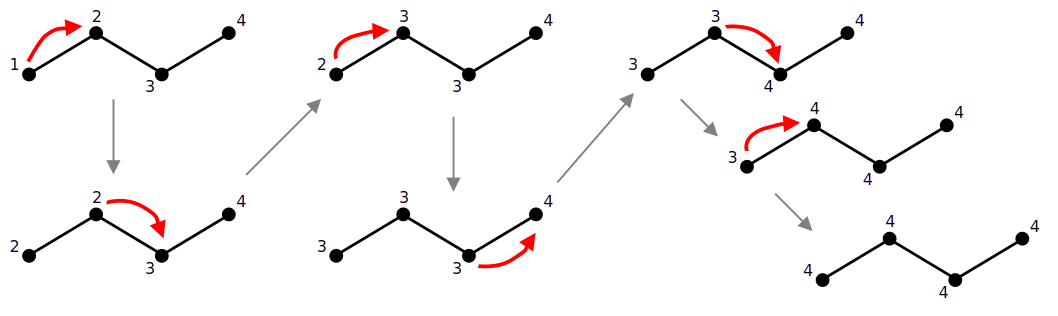}
			\caption{Chain cluster, where clustering edges share end vertices. Unfortunate winner assignment might make the clustering complexity explode.} \label{sec:algos-fig:chain_cluster}
		\end{figure}

		Our implementation of the graph pooling is the Treclus algorithm, which implements a threshold-based clustering. It creates a matching of the graph, where matched nodes are linked by an edge shorter than a threshold $\varepsilon$, and clusters matched nodes together. A graph matching of a graph $G=\{V, E\}$ is a subset of edges $E_m \subseteq E$ such that no two edges in $E_m$ share an end vertex \cite{lovasz_09}. It is very important that the clustering edges make a matching to keep a linear complexity of the clustering algorithm. Indeed, let us consider the cluster illustrated in Figure \ref{sec:algos-fig:chain_cluster}.
		Clusters are labelled by the highest node label they contain; for simplicity, assume the labelling procedure processes nodes 1 to 4 in order.
		Node 1 gets assigned cluster label 2. Then node 2 gets cluster label 3, but all nodes labelled as cluster 2 must be relabelled to cluster 3 as well. Finally, node 3 is labelled as cluster 4, and we must update the cluster labels of nodes 1 and 2. Any order of cluster labelling results in a similar case of non-linear complexity, and any attempt at optimising the labelling order loses generality and symmetry. This problem is however avoided if clusters can only contain two nodes, ensuring that clustering from a matching has a linear complexity. While we ideally wish to find a maximum matching of the subgraph $G' = \{V, E'=\left\{e\in E \left\vert e\leq \varepsilon\right\}\right\}$, that is the matching that maximises the number of matching edges, maximum graph matching is usually a quadratic problem \cite{behnezhad_24}, which is prohibitive in resource constrained environments. Note that the incomplete matching shortcoming can be solved by successive runs of the Treclus algorithm. Indeed, by deleting the edges between pairs of nodes clustered together, we may ensure that the repeated clustering performs a unique matching at every repeated step. This ensures linearity while still pooling most short edges ($<\varepsilon$). From each pooled node pair, one is selected randomly for the cluster position.
		
		The reduction step consists of a feature wise pooling of nodes within the same clusters. The choice of pooling operation for classification follows the same logic as the convolution-step aggregator (max). For energy regression, however, the reduction step is different: rather than mean, the aggregator must be a feature-wise sum, in order to preserve information about cluster size; a mean would make a cluster of 2 identical nodes indistinguishable from a cluster of 10.
		\begin{figure}
			\centering
			\includegraphics[width=0.8\textwidth]{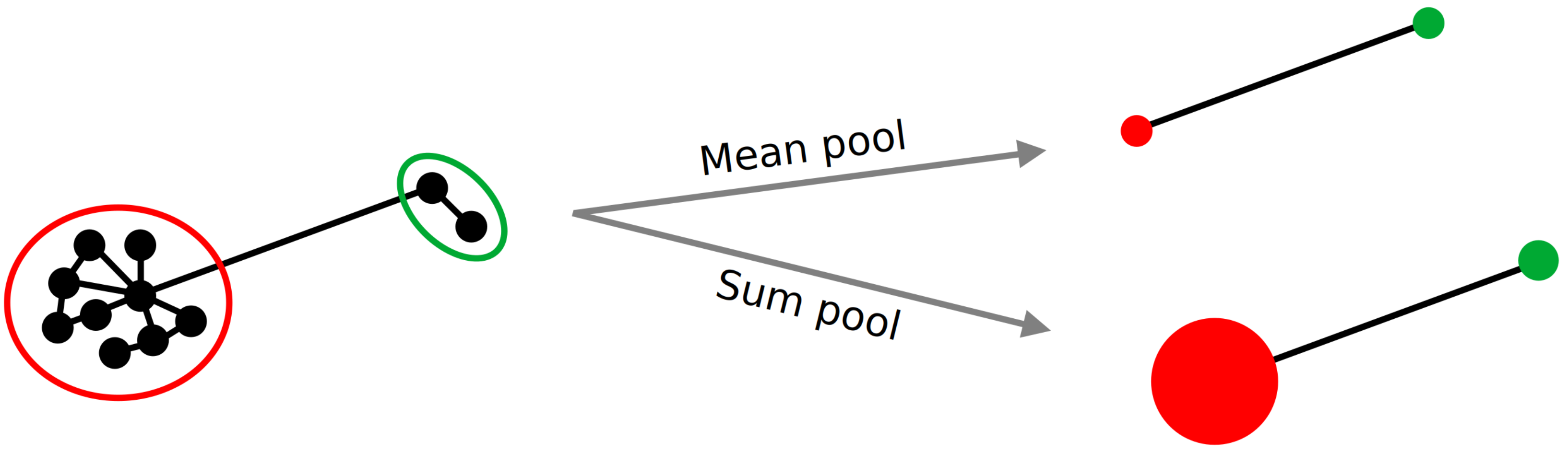}
			\caption{Using a sum aggregator for the reduction of large clusters ensure that node density information is maintained by the aggregator.}
		\end{figure}
		
		A last readout layer takes the learnt node representations and pools them across the full graph to extract the relevant graph-wide features. This output is then fed into a multi-layer perceptron (MLP) that will return the regression objective from the graph-wide feature representation.
		
	\section{Prompt Energy Regression}
				
		From the prompt signal, we are interested in recovering the energy of the prompt positron. Since most of the neutrino’s energy is passed on to the positron, we can use the rough estimate
		\begin{equation}
			\left\langle E_{e^+} \right\rangle \approx E_{\bar{\nu}_e} + m_p - m_n = E_{\bar{\nu}_e} - 1.293 \text{ MeV},
		\end{equation}
		for the incident neutrino’s energy. Here, we present a GNN pipeline to regress the energy of the positron.
		
		\subsection{Signal Pre-Treatment}
\begin{figure}
				\centering
				\includegraphics[width=0.55\textwidth]{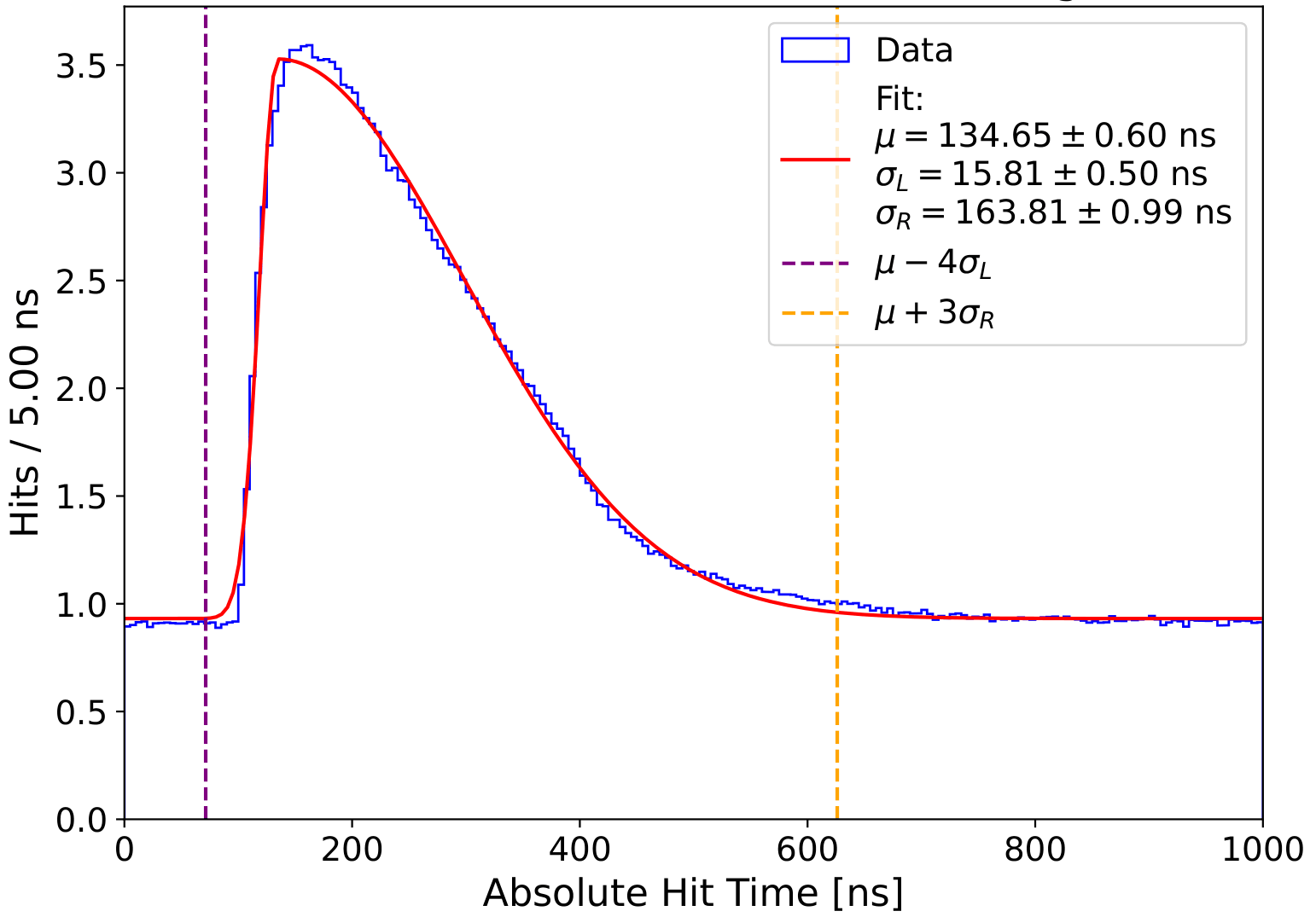}
				\caption{Average hit timings in $e^+$ events with energies in $[3, 30]$ MeV. Fitting the signal with an asymmetric Gaussian and a constant background term, we obtain two critical timings around the maximum when the signal is competitive with the background at $\mu^{-4\sigma_\text{L}}_{+3\sigma_\text{R}}$.}\label{sec:prompt-subsec:pre_process-fig:time_fit}
			\end{figure}

			In the previously presented detector geometry, a positron signal will trigger 50 to 200 PMT signals, dependent on its energy. If we look at a 40 $\mu$s window around the signal that is usually kept for analysis \cite{abe_26}, we end up with roughly $10^4$ hits. This is not suitable for GNN pipelines, first of all because the signal will be completely drowned under dark noise, but most importantly because this is too big for graph construction in a reasonable time. We thus need to restrict the signal down to only the positron induced hits. To do so, we first look at the average time distribution of hits in an event, and observe that the prompt hits dominate the dark noise. Fitting the signal with an asymmetric Gaussian, we can extract the critical times when the positron signal is competitive with the background, at $\mu^{-4\sigma_\text{L}}_{+3\sigma_\text{R}}$, as shown in Figure \ref{sec:prompt-subsec:pre_process-fig:time_fit}. The choice of $4\sigma_\text{L}$ for the lower bound was made to account for the small difference between the fitted mean $\mu$ and the maximum of the function, which we will use as a proxy for the mean.
			By sliding a 5 ns window on the timing distribution of hits in a prompt event, we identify the peak time $t_\text{max}$ of the timing distribution. We then select a window $[t_\text{max}-4 \sigma_\text{L}, t_\text{max}+3 \sigma_\text{R}]$, where the regressed values for the boundaries are 
			\begin{gather*}
				\sigma_\text{L} = 15.81 \text{ ns}, \\
				\sigma_\text{R} = 163.81 \text{ ns}.
			\end{gather*}
			This filtering approach leaves us with, on average, 200 nodes per graph, $46\%$ of which are from dark noise hits, which corresponds to $44\%$ of the total charge, leaving us with reasonably sized events for graph construction.
			
			Graphs are constructed using the $k$ nearest neighbours algorithm on the node positions in space-time cylindrical coordinates. The resulting graph has nodes $V$ with features: $(r_v, \cos\theta_v, \sin\theta_v, z_v, \tau_v=t_v-t_\text{max}, q_v)$, with $(r_v, \theta_v, z_v)$ the cylindrical coordinates of hits and $q_v$ the charge in photo-electrons (p.e.). Note the change of coordinate system from euclidean to cylindrical, which was done to respect the detector's geometry. Furthermore, we use coordinates $\cos\theta$ and $\sin\theta$ as they remove the discontinuity at $\theta=0$. Edges $E$ have as feature the space-time distance between the nodes.
			
		\subsection{Energy Regression Model}
						
			The energy regression model is based on the previously presented GNN structure. The first layer is a normalisation layer, which takes the raw input and rescales it following
			\begin{gather*}
				\tau_v\mapsto \frac{\tau_v - \mu_t}{\sigma_t}, \qquad q_v \mapsto \frac{q_v-\mu_q}{\sigma_q}, \\
				r_v \mapsto \frac{r_v}{R}, \qquad z_v \mapsto \frac{z_v}{H_{1/2}}, \\
				e_{vw} \mapsto \frac{e_{vw} - \mu_e}{\sigma_e},
			\end{gather*}
			where the average and standard deviation values are taken from the training set, $R$ is the radius of the HK cylinder, and $H_{1/2}$ the half height of the cylinder. The normalised graphs are passed through successive layers of convolution and pooling layers, using the Treclus pooling algorithm, and implementing concatenating skip connections to smoothen the loss landscape \cite{huang_17}. The convolution uses a mean as aggregator, and we use the sum for the pooling reduction operator. The node features are then pooled using a feature-wise sum to obtain a graph-wide representation. To avoid any outlying convolution outputs, mostly due to the sum poolings, we apply a batch norm to this readout output. Lastly, we apply a MLP, and rescale the outputs to the expected range, using
			\begin{equation*}
				E_\text{out} \mapsto 27E_\text{out} + 3.
			\end{equation*}
			
			The model was trained using an Adam optimiser \cite{kingma_17}, with learning rate $10^{-4}$, and an MSE loss function. The trained model had $\sim 5 \times 10^5$ parameters.
			
		\subsection{Performance}
			
\begin{figure}
				\centering
				\includegraphics[width=0.9\textwidth]{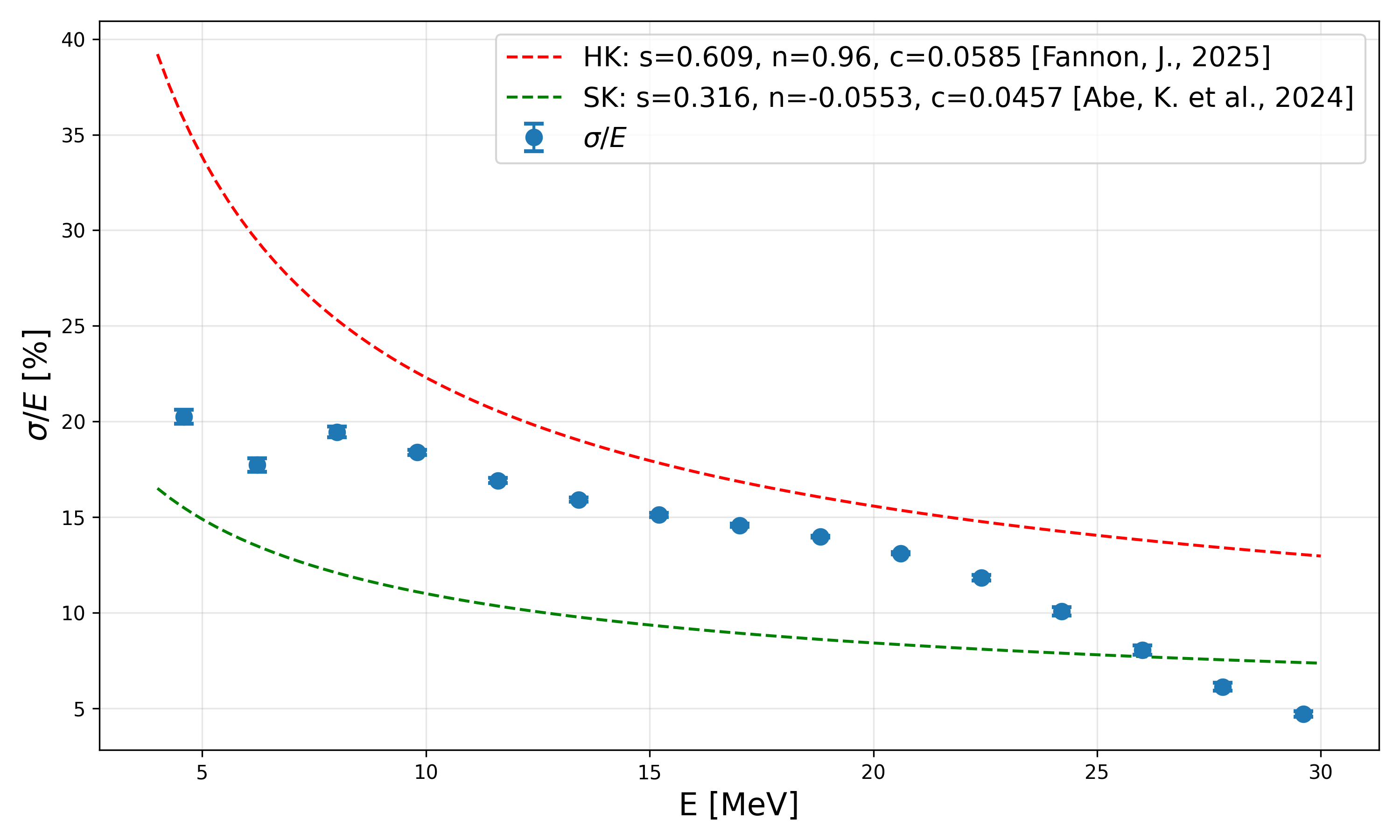}
				\caption{Energy resolution of the model, compared to the performance of energy regression using the effective number of hits $N_\text{eff}$ in Super and Hyper-Kamiokande.}\label{sec:prompt-subsec:energy-fig:resolution}
			\end{figure}

			We look at the absolute errors in energy $\Delta E = E_\text{pred} - E_\text{true}$, and define the model's resolution as the ratio $\sigma_{\Delta E} / E_\text{true}$. The resolutions of the model dependent on the positron energies are shown in Figure \ref{sec:prompt-subsec:energy-fig:resolution}.
			We compare the performance of our model with the resolutions obtained using the effective number of hits $N_\text{eff}$ \cite{abe_11}, computed by correcting the net number of hits with relevant extracted variables, which is the method used for energy regression in Super-Kamiokande \cite{abe_24}, and which was extended to Hyper-Kamiokande \cite{fannon_25}. We observe that we have a resolution that is comparable to that in Super-Kamiokande, which is notable given the lower PMT coverage and the increased absorption and scattering associated with the larger detector volume. This comparison should however be interpreted as a reference to published low-energy reconstruction performance rather than as a direct benchmark, since the detector configurations and reconstruction conditions are not identical.
			
			It is interesting to observe that the resolution curve does not follow the typical $\sigma(E)/E = s/\sqrt{E} \oplus n/E \oplus c$ \cite{abe_24, fannon_25}, with $\oplus$ the quadratic sum. One possible interpretation is that this is due to the fully data-driven training of the algorithm. At lower energies, where the events leave only a few hits, the model might regresses the energy by identifying the deviation from raw noise. This would explain the relatively good performance in the lower energy ranges. In the higher energy ranges, the model might observe clear patterns, which would explain the drop in resolution from 20 MeV. In the middle of the energy range, the model possibly struggles with the irregular signatures of Cherenkov cones at the boundary between noisy and well defined. 
			
			Ways by which physics knowledge can be implemented in the pipeline include, first of all, the inclusion of $N_\text{eff}$ in the model, as a graph-wide feature, or as a pre-selection criterion to use one of two models, one for high and one for low energy candidates. It may also be used in the loss term, as a way to have two terms, one for the low $N_\text{eff}$/energy values, and one for the high $N_\text{eff}$/energy range.
		
	\section{Neutron Tagging}\label{sec:neutron_tagging}
		
		Once a Cherenkov Signal is identified as a potential IBD prompt signal, one looks in the following 500 $\mu$s for a signal corresponding to the emission of a photon after a neutron capture \cite{smy_07, elhedri_21}. This is a most difficult task, as the neutron capture emits a $\gamma$ with a 2.2 MeV energy, which results, on average, in 7 hits in the detector, drowned in 500 $\mu$s of dark noise, producing $10^5$ hits. In this section, we introduce our neutron tagging pipeline.
		
		\subsection{Signal Pre-Processing}
		
\begin{figure}
			\centering
			\begin{subfigure}{0.48\textwidth}
				\centering
				\includegraphics[width=\textwidth]{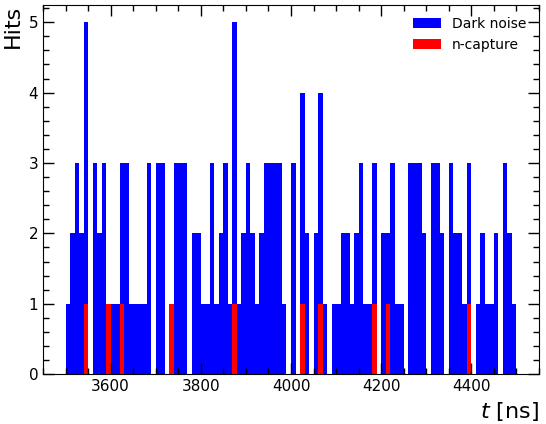}
				\caption{Hit timings in detector time.}
			\end{subfigure}
			\begin{subfigure}{0.48\textwidth}
				\centering
				\includegraphics[width=\textwidth]{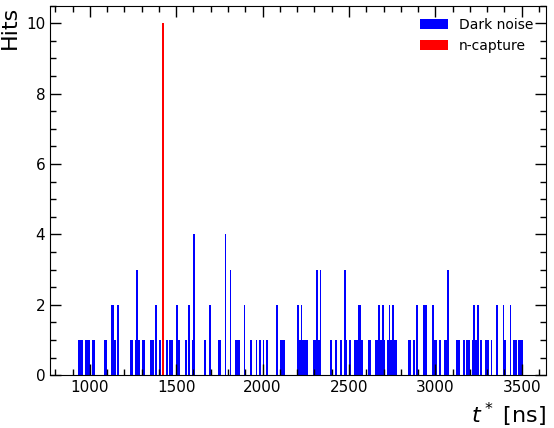}
				\caption{Hit timings in TOF-corrected time space.}
			\end{subfigure}
			\caption{Taking the PMT hit timings from the detector space (a) to the TOF-corrected space (b) concentrates the neutron hits within a 10 ns window.}\label{sec:n_tag-subsec:pre_process-fig:tof_correction}
		\end{figure}

\begin{figure}
			\centering
			\includegraphics[width=0.5\textwidth]{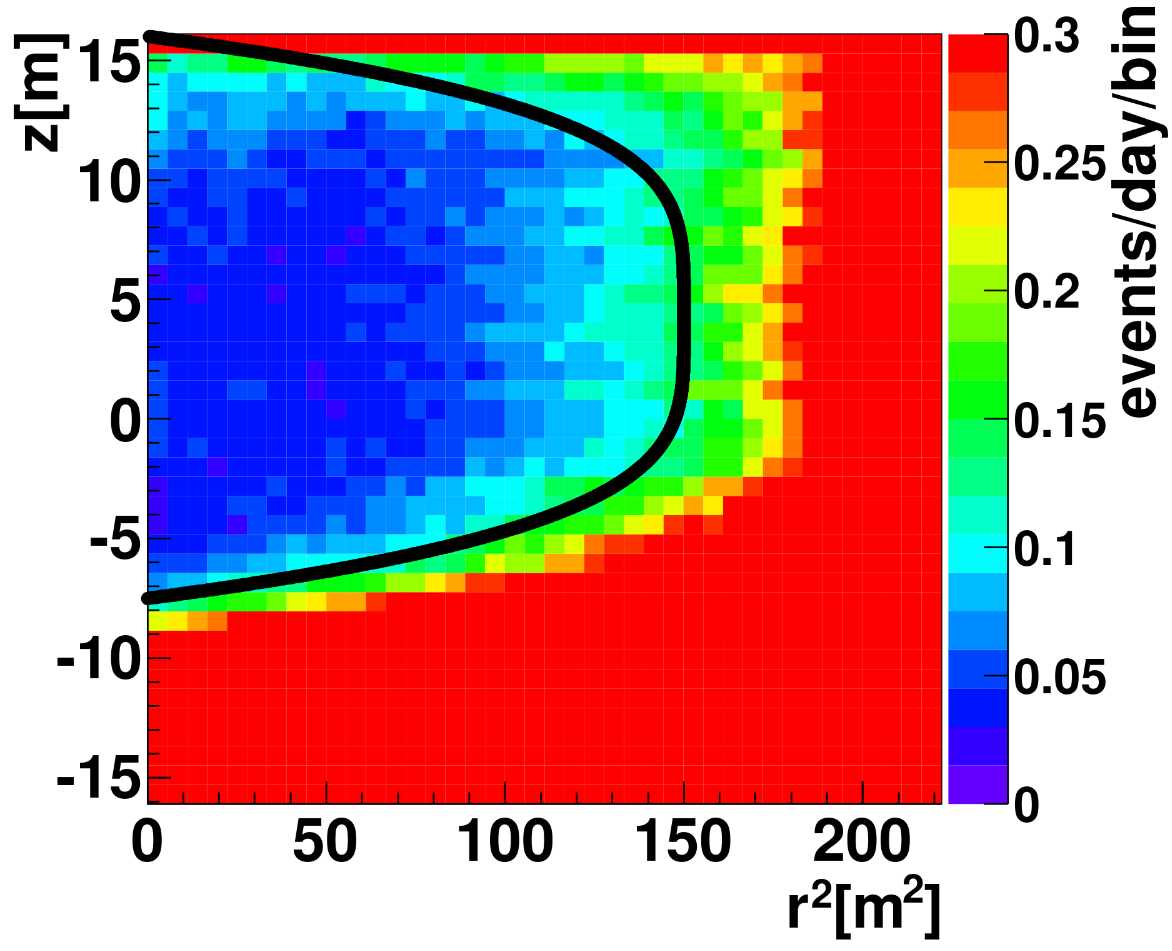}
			\caption{Radon decay frequencies in the Super-Kamiokande geometry. From \cite{abe_16}}\label{sec:n_tag-subsec:pre_process-fig:sk_rn_decays}
		\end{figure}

		Since the signal is drowned in a sea of dark noise hits, one must first extract a subset of hits which includes as many of the true neutron hits as possible. Based on the pre-selection algorithm used in Super-Kamiokande \cite{giampaolo_23}, we use the previously regressed vertex for the IBD interaction, and use it as a first estimate for the neutron capture and subsequent photon emission. As such, correcting the timing of hits by the time of flight (TOF) of a photon from the vertex to the node's coordinates, while the background will remain uniform, one should obtain all true neutron hits at the exact same moment, corresponding to the timing of the neutron capture, as illustrated in Figure \ref{sec:n_tag-subsec:pre_process-fig:tof_correction}.
		An issue with this approach is that it does not take into account the errors in vertex position and timing estimation. As such, the spread of the true neutron hits in the time of flight corrected space is of roughly 10 ns, taking into account the projected resolution for HK \cite{hk_18}. Note that since this approach also spreads the dark noise hits, as the previously uniform time distribution now includes the variance in TOF from vertex to dark noise hit. This effectively reduces the dark rate. Another advantage of the candidate selection method is that it reduces the frequency of accidental coincidences. Indeed, in the case of accidental coincidences, the chance that the vertices for the prompt trigger and the coincidental signal be the same are very low, and correcting $t-\tilde{t}$ will spread the coincidental hits over a large time window, not allowing for a candidate to be formed. In particular, we look at radon decays. Decays of $\ce{^{222}Rn}$ from the rocks surrounding the Hyper-Kamiokande cavern will produce $\ce{^{214}Bi}$ through its daughter chain, which will $\beta$ decay producing an electron of energy $\sim 3.27$ MeV \cite{abe_16}, which competes with the neutron capture signal at $2.2$ MeV. Based on the analysis of the frequency of radon decays in Super Kamiokande (Figure \ref{sec:n_tag-subsec:pre_process-fig:sk_rn_decays}), we extrapolate it to Hyper-Kamiokande, assuming the same frequency of radioactive events in the same fraction of HK volume, which is a pessimistic evaluation. We find that a radon event happens in a delayed signal about once every 60 events. If we take into account the fact that the radon event vertex must coincide with the prompt vertex, radon events qualify as neutron candidate at a frequency of less than $10^{-5}$ and can thus be considered negligible.
		
\begin{figure}
			\centering
			\includegraphics[width=0.6\textwidth]{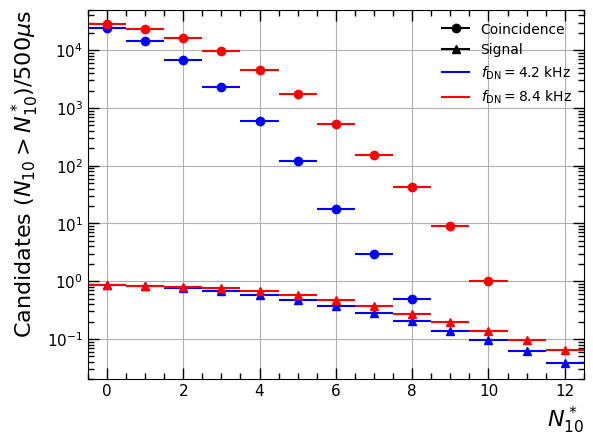}
			\caption{Candidate selection frequency for signal and dark noise events, depending on $N^*_{10}$. The trade-off signal to noise ratio was selected for $N_{10}^* = 7$.}\label{chap:dsnb-sec:n_tag-subsec:pre_process-fig:n10}
		\end{figure}

		Still using the Super-Kamiokande approach \cite{giampaolo_23}, once the time coordinate has undergone the TOF correction $t\mapsto t^*$, we use a 10 ns sliding window on the hit time distribution to select candidates for neutron hits cluster. We define a threshold $N^*_{10}$, which is the minimum hit count $N_{10}$ needed within a 10 ns window to be considered a candidate sub-event. This threshold was selected according to the data presented in Figure \ref{chap:dsnb-sec:n_tag-subsec:pre_process-fig:n10}.
		The threshold for a sufficient signal to noise ratio was selected for $N^*_{10} = 7$, so we select, for each event, the 10 ns subsets respecting $N_{10} > N^*_{10}$. This results in $\sim 300$ candidates per 500 $\mu$s delayed signal search window.
		
		Once neutron candidates have been selected, we use a space-time $k$-NN algorithm, using $k=4$, to obtain a graph where nodes have as feature the PMT charge $q$, and edge features are the space-time distance between the nodes. Again, the node features are $(r_v, \cos\theta_v, \sin\theta_v, z_v, \tau_v=t_v-\min(t_v), q_v)$, where the timing shift is done with respect to the first recorded hit of the candidate. The edges of the graph have the space-time length as single feature.
		
		\subsection{Model and Training}
		
		The model we use for this task is once again the one presented for the prompt signal energy regression task. We use a maximum aggregator for the convolution and readout operations. We also do not implement any pooling, as the average graph diameter (the longest of all shortest paths between any two nodes) for graphs with $\sim 10$ nodes and built with $k=4$ is equal to 2.5, which is much less than the depth of the convolution network. Indeed, this last statement implies that all information can already flow through the graph, and as such, additional pooling is therefore not required for graph-wide information propagation and was omitted to avoid unnecessary coarsening.
		
		The full pipeline for neutron tagging therefore takes the graphs obtained from neutron candidates, and passes them through several layers of convolution, followed by a readout pooling layer, taking the maximal features across all nodes of the graph. This output is then passed through an artificial neural network to output the binary class probabilities for either dark noise or signal. A softmax is applied to this output to ensure probabilities which sum up to 1. The model is trained using the Categorical Cross Entropy Loss. 
		
		It is important to note that to perform a training where the model may identify patterns corresponding to a 2.2 MeV $\gamma$ emission, one must impose a minimal number of true neutron hits to be considered as a true neutron candidate. For the training, we therefore only label as true neutron candidates those that contain at least 4 true neutron hits, which we consider to be enough to be characteristic of a Cherenkov signal. 
		
		\subsection{Performance}
		
		This model is a bare-bones approach to neutron tagging in a Hyper-Kamiokande like detector. As such, it does not implement any fiducial volume or pre-selection algorithm that a fully deployed model might. We therefore propose it as a rough baseline for what can be done using graph convolution models towards the goal of neutron tagging and DSNB detection in water-Cherenkov detectors, for neutron candidates satisfying the selection described above.
		
\begin{figure}
			\centering
			\includegraphics[width=0.6\textwidth]{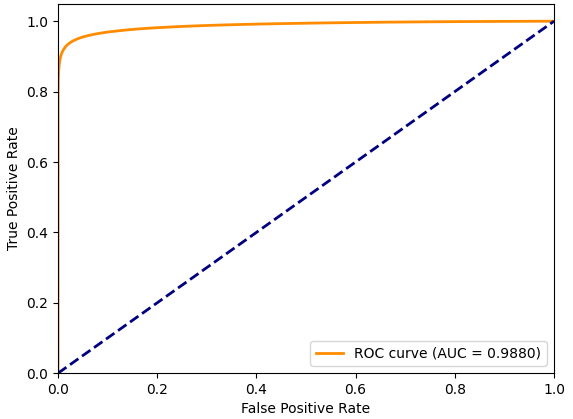}
			\caption{Receiver Operating Characteristic (ROC) curve for the Neutron tagging pipeline. The area under the curve (AUC) is a good performance metric in a context where we wish to reduce background contamination, such as DSNB detection.}\label{sec:n_tag-subsec:results-fig:roc}
		\end{figure}

		We first look at the Receiver Operating Characteristic (ROC) curve, represented in Figure \ref{sec:n_tag-subsec:results-fig:roc}.
		Choosing the class based on the highest predicted probability, we obtain the confusion matrix presented in Table \ref{sec:n_tag-subsec:results-tab:confusion_matrix}.
		\begin{table}
			\centering
			\begin{tabular}{cc||cc}
				\multicolumn{2}{c||}{\multirow{2}{*}{}}                   & \multicolumn{2}{c}{Predicted} \\ \cline{3-4} 
				\multicolumn{2}{c||}{}                                    & Coincidence   & Neutron       \\ \hline\hline
				\multicolumn{1}{c|}{\multirow{2}{*}{True}} & Coincidence  & 97.47 \%      & 2.53 \%      \\
				\multicolumn{1}{c|}{}                      & Neutron      & 6.02 \%       & 93.98 \%
			\end{tabular}
			\caption{Confusion matrix for the Neutron tagging model.}\label{sec:n_tag-subsec:results-tab:confusion_matrix}
		\end{table}
		The classifier correctly rejects 97.47\% of background candidates while identifying 93.98\% of the selected neutron candidates. The asymmetric error rates are favourable in a background-dominated search.
		
		The overall performance of the model could be improved by adding more of the detector's specifications in the neutron tagging pipeline. In particular, the prompt signal regression performance could have a major impact in the candidate pre-selection by reducing the candidate selection window size, and thus improving the candidate purity rate. Imposing a fiducial volume filtering could also improve the model, by improving the quality of the vertex regression, as well as improving the quality of the Cherenkov signal. It would also further reduce the radon contamination of the signal, radon disintegration mostly happening close to the cavern walls, i.e. closer to the detector's sides.
	
	\section{Conclusion}
		
		In this work, we have shown that GNNs provide a promising framework for the varied tasks of low energy event reconstruction in Water-Cherenkov detectors. We have presented a message passing convolution based pipeline for reconstruction from raw hits of low energy events in an HK-like detector, and applied it to energy regression of IBD prompt signals and neutron tagging.
		
		For the regression of the energy of a positron in the [3, 30] MeV range, we have recovered a resolution that is comparable to the state of the art. In particular, we observe increased performance for $E_{e^+} > 20$ MeV. This is explained from the better signal definition at higher energies. For the neutron tagging, we obtain a good classification performance. In particular, we observe that our classifier has a tendency to reject undefined neutron signals, that are similar to noise. This is a wanted feature, as while we are interested in detecting as many neutrons as possible, a high signal purity is a key requirement to a successful neutron tagging model. We propose this classifier's performance as a baseline for the application of modern deep learning architectures to the problem of neutron tagging in next generation water-Cherenkov detectors.
		
		The above results indicate that GNN-based reconstruction is most naturally integrated with established physics-driven reconstruction methods rather than used as a fully stand-alone approach. As the signals of Cherenkov light are projections of Cherenkov cones on an irregular and anisotropic geometry. This requires further work on the raw data before a GNN can be fully exploited. Using previously existing frameworks is essential to best exploit the detector's capacities. For instance, one could think of using a GNN architecture to validate potential IBD vertices for grid-search based algorithms, such as the BONSAI algorithm \cite{smy_07}. For the energy regression of the prompt positron, pre-existing methods adapt the energy estimate to the effective number of recorded hits. One could imagine improving the performance over the full energy range by training multiple models for specific ranges of the effective hit count.
	
	\section*{Acknowledgments}
	
		The authors are thankful to the ANR for the funding of the OGCID project, of which this study is part of, through funding ANR-21-CE31-0030.

	\printbibliography
	
\end{document}